# Non-Parametric Time Between Events and Amplitude Methods for Monitoring Drought Characteristics

AUTHOR
Michele Scagliarini ORCID: 0000-0003-3209-4745
Department of Statistical Sciences, University of Bologna, Italy
Via delle Belle Arti, 41, 40126, Bologna, Italy
Email: michele.scagliarini@unibo.it

**Abstract**

Drought is a significant natural phenomenon with profound environmental, economic, and societal impacts. Effective monitoring of drought characteristics -such as intensity, magnitude, and duration- is crucial for resilience and mitigation strategies. This study proposes the use of non-parametric Time Between Events and Amplitude (TBEA) control charts for detecting changes in drought characteristics, specifically applying them to the Standardized Precipitation and Evapotranspiration Index. Aware of being non-exhaustive, we considered two non-parametric change-point control charts based on the Mann-Whitney and Kolmogorov-Smirnov statistics, respectively. We studied the in-control statistical performances of the change-point control charts in the time between events and amplitude framework through a simulation study. Furthermore, we assessed the coherence of the results obtained with a distribution-free upper sided Exponentially Weighted Moving Average control chart specifically designed for monitoring TBEA data. The findings suggest that the proposed methods may serve as valuable tools for climate resilience planning and water resource management.

# 1 Introduction

Drought is a significant natural phenomenon with the potential to cause profound and far-reaching impacts on the agricultural production, the environment, the economy and social stability. Therefore, it is important to properly assess this phenomenon and objectively quantify the characteristics of drought episodes in terms of intensity, magnitude and duration.

The Standardized Precipitation and Evapotranspiration Index, SPEI, has been proposed for this purpose (Vincente-Serrano et al., 2010). The SPEI is a climatic drought index based on precipitation, temperature and potential evapotranspiration. It is used to assess drought characteristics and identify the onset and end of drought episodes. SPEI allows for comparison of drought severity over time and space, as it can be calculated over a wide range of climates and accumulation periods ranging from 1 to 48 months (referred to as SPEI-1, SPEI-2, and so on). These different timescales are used to reflect the impact of drought on different water-related sectors:

- Meteorological and soil moisture conditions (agriculture drought) respond to precipitation and temperature anomalies on relatively short time scales, such as 1-6 months (SPEI-1 to SPEI-6);
- River flow, reservoirs, and groundwater (hydrological drought) respond to longer-term anomalies of the order of 6 months to 12 months or longer (SPEI-6 to SPEI-12);
- Situations in which the water supply fails to satisfy water demand, resulting in negative consequences for society, the economy, and the environment are referred as socio-economic droughts. For this last case the appropriate timescale is 24 months (SPEI-24).

The standardized precipitation and evapotranspiration index can assume positive and negative values with the following moisture categories (Liu et al., 2024):

- $SPEI \geq 2$ extremely wet conditions
- $1.5 \leq SPEI < 2$ very wet conditions
- $1 \leq SPEI < 1.5$ moderately wet conditions
- $-1 < SPEI < 1$ normal conditions
- $-1.5 < SPEI \leq -1$ moderately dry conditions
- $-2 < SPEI \leq -1.5$ severely dry conditions
- $SPEI \leq -2$ extremely dry conditions.

To design effective drought management and resilience policies, it is crucial to monitor drought evolution and detect changes in drought characteristics. Since the key drought characteristics are severity, duration and frequency, a suitable methodology for monitoring drought events should consider the time interval ($T$) between two occurrences and the magnitude ($X$) of each event.

Time Between Events and Amplitude (TBEA) control charts have been proposed to monitor this type of phenomenon: a decrease in $T$ and/or an increase in $X$ may result in a negative condition that needs to be monitored and possibly detected with control charts.

This research proposes using distribution-free TBEA control charts to detect changes in drought characteristics. A non-parametric approach was selected due to the challenges in determining the underlying distributions of both the time between events and the event amplitudes. Specifically, and without claiming exhaustiveness, we focused on two non-parametric change-point control charts based on the Mann-Whitney (MW) and Kolmogorov-Smirnov (KS) statistics, respectively (Hawkins and Deng, 2010; Ross and Adams, 2012). Through a case study and simulation experiment, the statistical properties of these change point control charts are evaluated and compared with those of an Exponentially Weighted Moving Average (EWMA) control chart specifically designed for monitoring TBEA data (Wu et al. (2021).

To the best of our knowledge, this is the first study to propose distribution-free change-point control charts for monitoring TBEA data. We find this approach particularly promising because, beyond merely signaling an alarm (i.e., identifying the detection time), it can also provide an estimate of the time at which the change occurred (i.e., the change-point). Having such an estimate enhances the overall understanding of the monitored process:

- in environmental applications, it enables a deeper interpretation of the underlying dynamics of the phenomenon;
- in industrial contexts, it supports more effective tracing of potential assignable causes.

The paper is structured as follows. Section 2 details the distribution-free control charts considered for our purpose. Section 3 introduces the data and presents the main results. Section 4 compares and discusses the results. Section 5 presents further developments in which a different accumulation period for drought assessment is considered. Finally, Section 6 provides the concluding remarks.

**2 Methodology**

In real practice it is often of interest to monitor the occurrence of undesirable events such as equipment failure, quality problems or extreme natural phenomena. In these situations, two characteristics are usually relevant: the time interval $T$ between two occurrences of the negative event $E$ and the associated magnitude $X$.

Time Between Events and Amplitude control charts have been specifically proposed to simultaneously monitor the time interval $T$ between successive occurrences and the magnitude $X$ of an event. The first contribution in this field is due to Wu et al. (2009), who proposed a charting procedure to monitor the ratio $X/T$.

Generally, TBEA control charts use, as statistics to be monitored, some functions of the random variables $T$ and $X$,

$$Z = Z(T, X), \tag{1}$$

satisfying the following two properties (Castagliola et al. 2022):

$$\begin{aligned} & Z \text{ increases if either } T \text{ decreases or } X \text{ increases} \\ & Z \text{ decreases if either } T \text{ increases or } X \text{ decreases.} \end{aligned} \tag{2}$$

Most of the contributions in literature study the statistical properties of TBEA control charts under the assumption that the random variables $T$ and $X$ follow some specific distributions. To name but a few: Qu et al. (2013), Cheng et al. (2017), Ali and Pievatolo (2018), Qu et al. (2018), Rahali et al. 2019, Sanusi et al. (2020) and Rahali et al. (2021). However, in practice it is very difficult to identify the actual distribution of these variables. To overcome this difficulty, several TBEA distribution-free control charts have been proposed. Among the most recent ones we quote: Mukherjee et al. (2018), who introduced a distribution-free TBEA control chart for monitoring quality service data; Huang et al. (2018), who developed two non-parametric EWMA control charts; Wu at al. (2021), who proposed a non-parametric EWMA TBEA control chart based on sign statistics; He et al. (2022), who designed a non-parametric CUSUM scheme for monitoring multivariate time between events and amplitude data.

### 3.1 Distribution-free exponentially weighted moving average TBEA control chart

Let $T_i$ ($i=1,2,\ldots$) be the time intervals between two consecutive occurrences of a given event $E$ and let $X_i$ be the corresponding magnitudes of the event of interest. To monitor $T_i$ and $X_i$ simultaneously Wu et al. (2021) assumed that $T_i$ and $X_i$ are continuous random variables, both defined on $[0;+\infty)$, with unknown distribution functions $F_T(t|\theta_T)$ and $F_X(x|\theta_X)$ respectively, where $\theta_T$ and $\theta_X$ are known $\alpha$-quantiles. Without loss of generality, it is considered that $\theta_T$ and $\theta_X$ are the median values of $T_i$ and $X_i$, respectively, and when the process is in-control it follows that $\theta_T = \theta_{T_0}$ and $\theta_X = \theta_{X_0}$. Let us now introduce the sign statistics

$$ST_i = \text{sign}(T_i - \theta_{T_0}) \tag{3}$$

and

$$SX_i = \text{sign}(X_i - \theta_{X_0}), \tag{4}$$

for $i=1,2,\ldots$, where $\text{sign}(x) = -1$ if $x<0$ and $\text{sign}(x) = +1$ if $x>0$.

Therefore, the statistic

$$S_i = \frac{(SX_i - ST_i)}{2} \qquad (i=1,2,\ldots), \qquad (5)$$

has the following behaviour:

- $S_i = -1$ when $T_i$ increases and, at the same time, $X_i$ decreases (positive situation);
- $S_i = +1$ when $T_i$ decreases and, at the same time, $X_i$ increases (negative situation);
- $S_i = 0$ when both $T_i$ and $X_i$ increase or when both $T_i$ and $X_i$ decrease (intermediate situation).

Note that $S_i$ is a discrete random variable; therefore, it is impossible to accurately compute the run length properties of the control chart using a Markov chain approach. To solve this problem, Wu et al. (2021) developed the "continuousify" method, which consists in defining an extra parameter $\sigma \in [0.1, 0.2]$ and to convert $S_i$ into a continuous random variable, $S_i^*$, defined as a mixture of three normal random variables $Y_{i,-1} \sim N(-1, \sigma)$, $Y_{i,0} \sim N(0, \sigma)$ and $Y_{i,+1} \sim N(1, \sigma)$, respectively. More precisely $S_i^*$ is defined as:

$$S_i^* = \begin{cases} S_i^* = Y_{i,-1} & \text{if } S_i = -1 \\ S_i^* = Y_{i,0} & \text{if } S_i = 0 \\ S_i^* = Y_{i,+1} & \text{if } S_i = +1 \end{cases}, \qquad (6)$$

Note that the parameter $\sigma$ has to be fixed. However, it has been demonstrated that this parameter has no impact on the performance of the control chart if it falls within the suggested range.
The upper-sided EWMA TBEA control chart uses the statistic

$$Z_i^* = \max\left(0, \lambda S_i^* + (1-\lambda) Z_{i-1}^*\right), \qquad (7)$$

with an upper asymptotic control limit given by

$$UCL = K\sqrt{\lambda(\sigma^2 + 0.5)/(2-\lambda)}, \qquad (8)$$

where $\lambda \in [0, 1]$ and $K > 0$ are the control chart parameters and the initial value is $Z_0^* = 0$. The optimal design parameters $\lambda$ and $K$, such as to guarantee the desired chart performance, are be obtained by studying the statistical properties of the control chart by means of a Markov chain approach. Full details can be found in Wu et al. (2021).

*3.2 The change point model approach*

To properly monitor time between events and amplitude data using change points control charts, it is important to consider that the variables $T$ and $X$ can have very different scales. Therefore, to avoid favoring one variable over the other, it is appropriate to define and use the normalized variables $T' = T/\mu_{T_0}$ and $X' = X/\mu_{X_0}$, where $\mu_{T_0}$ and $\mu_{X_0}$ are the in-control means of $T$ and $X$, respectively. To define a TBEA statistic $Z$, function of $T'$ and $X'$, and satisfying the properties reported in (2), several proposals exist in literature. For example, Rahali et al. 2019 studied the statistical performance of TBEA control charts considering the statistics $Z_D = X' - T'$, $Z_R = X'/T'$ and $Z_P = X' + 1/T'$, assuming known distribution functions for $T'$ and $X'$. Among the possible choices for the TBEA statistic to be monitored, we consider the ratio $Z_R = X'/T'$ and propose to monitor this statistic with non-parametric change-point control charts.

In the Change-Point (CP) model framework, it is assumed that the values of the monitored statistic $z_1, z_2,...$ are generated by the random variables $Z_1, Z_2,...$ with unknown distribution functions $F_1, F_2,...$, respectively. A change point occurs at instant $\tau$ when $F_\tau \neq F_{\tau+1}$, and it is usually assumed that the observations are independent and identically distributed between every pair of change points. Therefore, the distribution of the sequence can be described by the following model:

$$X_i \sim \begin{cases} F_0 & \text{if } 0 < i \leq \tau_1 \\ F_1 & \text{if } \tau_1 < i \leq \tau_2 \\ ... \\ F_k & \text{if } \tau_k < i \leq m \end{cases}, \quad (9)$$

where $0 < i < \tau_1 < \tau_2 < \cdots < \tau_k < m$ denote $k$ unknown change points. Within this framework, it is of interest to test $H_0$: $k=0$, versus $H_1$: $k \neq 0$, to estimate the number of change points ($k$) and to estimate their locations $\hat{\tau}_1, \hat{\tau}_2,...,\hat{\tau}_k$.

For testing the aforementioned hypothesis system, Hawkins and Deng (2010) proposed a change point control chart designed to detect changes in the location of the process distribution, based on the Mann-Whitney (MW) *U*-statistic:

$$U_{k,n} = \sum_{i=1}^{k} \sum_{j=k+1}^{n} D_{ij} \quad 1 \leq k \leq n-1, \tag{10}$$

where

$$D_{ij} = \text{sign}(Z_i - Z_j) = \begin{cases} 1 & \text{if } Z_i > Z_j \\ 0 & \text{if } Z_i = Z_j \\ -1 & \text{if } Z_i < Z_j \end{cases}. \tag{11}$$

The variance of $U_{k,n}$ depends on the split point $k$, so it is suitably standardized, thus obtaining the statistic $T_{k,n}$ (Hawkins and Deng 2010). Therefore, the test for the presence of a change point and the estimate of its time of occurrence are given by maximizing $T_{k,n}$ over $k$,

$$T_{\max,n} = \max_{1 \leq k \leq n-1} |T_{k,n}|. \tag{12}$$

Note that in this formulation, the data set size $n$ is no longer fixed but grows as long as the process is monitored and every instant of time is analysed as a potential change point by carrying out a two-sample Mann-Whitney test between the observations before and after that time. If $T_{\max,n}$ exceeds a specified control limit $h_{n,\alpha}$, the method signals that a change has occurred and estimates the time of change as the maximizing split point:

$$\hat{\tau}_T = \arg\max_{1 \leq k \leq n-1} |T_{k,n}|. \tag{13}$$

The thresholds $h_{n,\alpha}$ are such that the conditional probability of a false alarm at any $n$ is constant and equal to $\alpha$ and are established by simulation (Hawkins and Deng 2010). The MW control chart is implemented in the *R* package *cpm* (Ross, 2015).
To detect arbitrary changes in the process distribution, Ross and Adams (2012) proposed a control chart based on the Kolmogorov-Smirnov (KS) statistic. This control chart tests for a change point immediately following any observation $z_k$ by partitioning the observations into two samples, $S_1=(z_1,\ldots,z_k)$ and $S_2=(z_{k+1},\ldots,z_t)$, and subsequently comparing the corresponding empirical distribution functions $\hat{F}_{S_1}(z)$ and $\hat{F}_{S_2}(z)$.

The test statistic is based on the maximum difference between the empirical distributions

$$D_{k,t} = \sup_z \left| \hat{F}_{S_1}(z_i) - \hat{F}_{S_2}(z_i) \right|. \tag{14}$$

and a change point is detected if $D_{k,t}$ exceeds a specified threshold $_D h_{k,t}$.

Whenever a new observation $z_k$ is received, the sequence $(Z_1,\ldots,Z_t)$ is considered as being a fixed-size sample. If $H_0$ is rejected, the estimate $\hat{\tau}$ of the change point location is the value of $k$ that maximizes $D_{k,t}$. The thresholds $_D h_{k,t}$, such as to guarantee the desired $ARL_0$, are determined by means of the Monte Carlo simulation.

For the sake of completeness, it is worth noting that $D_{k,t}$ has variance which depend on $k$. Consequently, the test is performed on the standardized versions of the statistic $D_{k,t}$. Full details can be found in Ross and Adams (2012). Also the KS control charts is implemented in the *R* package *cpm* (Ross, 2015).

## 3. Data and Case Study

SPEI data are available from the Global SPEI database SPEIbase v2.9 (https://spei.csic.es/database.html) with time scales from 1 to 48 months, a spatial resolution of 0.5° lat/lon (approximately 55 km), and temporal coverage from January 1901 to December 2022 (Beguería et al. 2014).

Among the possible timescales for the standardized precipitation and evapotranspiration index, the 12-month accumulation period was chosen because SPEI-12 is a reliable measure of hydrological drought. The SPEI-12 data were downloaded from the Global SPEI database for the coordinates of Bologna (44.4949° N, 11.3426° E), where observations are available from December, 1901 to December, 2022 (1453 observations).

Figures 1 and 2 show the time series plot and the histogram, respectively, of the SPEI-12 data. The Lilliefors normality test ($D$=0.012044, *p-value*=0.8768) does not reject the hypothesis of normal distribution and Table 1 provides some summary statistics.

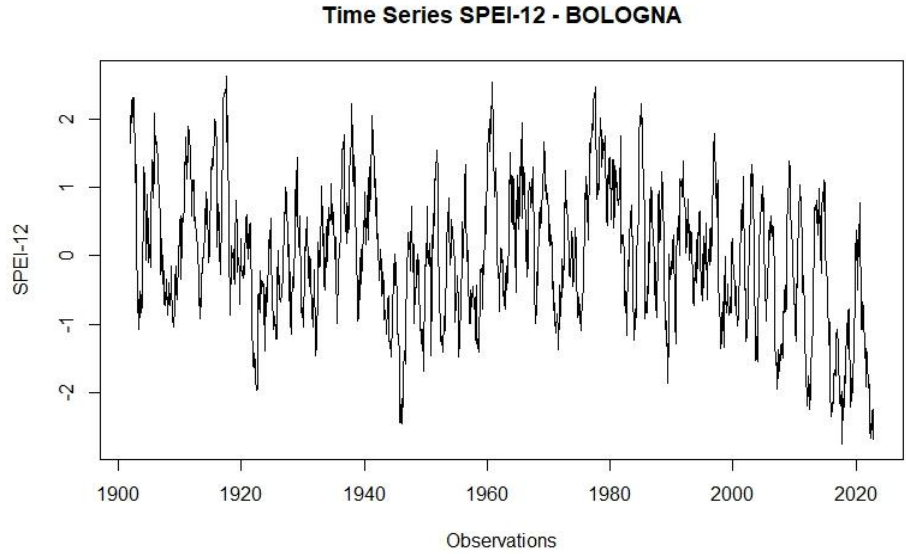

**Figure 1:** Time sequence plot of the available SPEI-12 data for the period December, 1901–December, 2022

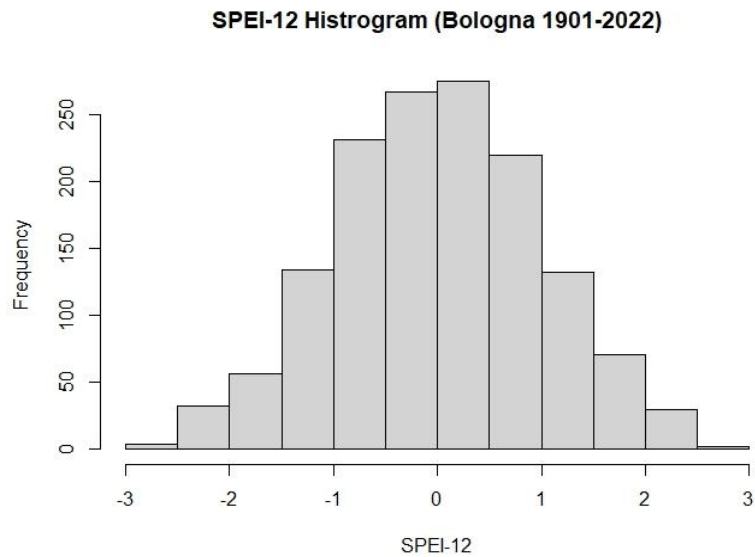

**Figure 2:** Histogram of the available SPEI-12 data

| Min. | First Qu. | Median | Mean | Third Qu. | Max. | Variance |
|---|---|---|---|---|---|---|
| -2.7563 | -0.7070 | 0.0069 | 0.0003 | 0.6789 | 2.6351 | 0.9891 |

**Table 1**. Descriptive statistics of SPEI-12 data

The aim is to monitor the characteristics of drought episodes; therefore, the occurrences of SPEI-12≤-1, corresponding to drought conditions, were considered as the events (*E*) of interest. As the TBEA methodology assumes that event amplitudes are defined on [0; +∞), the absolute values of SPEI-12 at the occurrence dates were considered as the corresponding amplitudes $X_i$ and, given the monthly frequency of the SPEI-12, the time $T_i$ between two events is computed in months.

The available dataset contains 226 events $E$ (values of SPEI-12≤-1) and Figures 3 and 4 illustrate the absolute values SPEI-12≤-1 ($X_i$) and the time ($T_i$) between the drought events $E$, respectively.

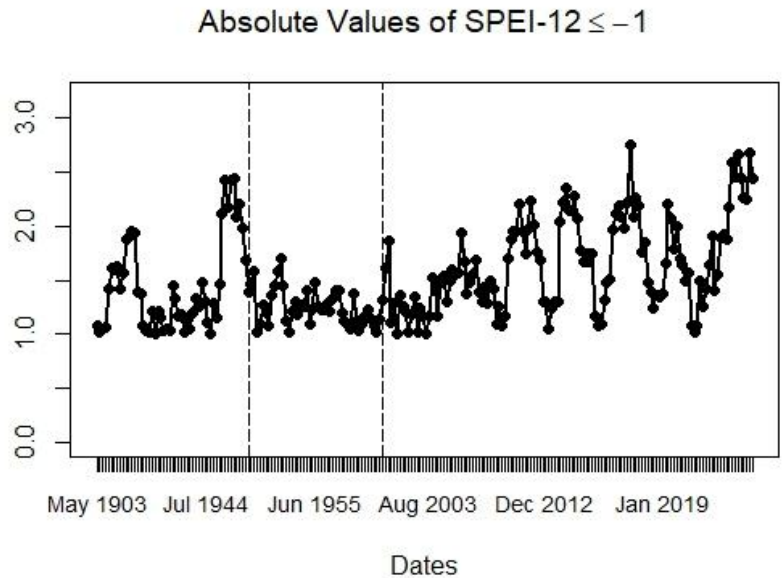

**Figure 3**: Amplitudes $X_i$ of the drought events $E$ (data for Phase-1 are indicated between dashed vertical lines)

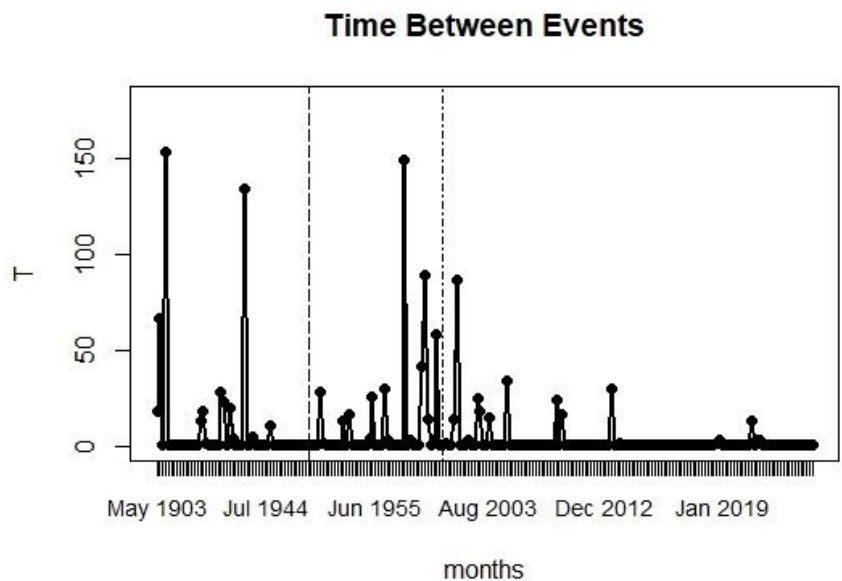

**Figure 4** Time ($T_i$) in months between drought events $E$ (data for Phase-1 are indicated between dashed vertical lines)

For performing the Phase-1 we selected the time period from July 1946 (observation #536) to July 1989 (observation #1052), comprising 517 SPEI-12 observations, with mean and the standard deviation $\mu_{SPEI-12} = 0.2082$ and $\sigma_{SPEI-12} = 0.9213$, respectively. In this period, marked by dashed vertical lines in Figures 3 and 4, occurred 47 events $E$ (values of *SPEI*-12≤-1), which were used to estimate in-control median (for the EWMA chart) and mean (for the CP charts) values: $\hat{\theta}_{T_0} = 1$, $\hat{\theta}_{X_0} = 1.2263$, $\hat{\mu}_{T_0} = 10.9149$ and $\hat{\mu}_{X_0} = 1.2511$, respectively.

To ensure comparability, all control charts were configured with the same in-control theoretical statistical properties (ARL$_0$=370 or with a theoretical false alarms rate equal to $\alpha_0 = 0.0027$).

## 3.1 EWMA-TBEA control chart

To obtain a control chart with an in-control ARL equal 370, parameters $\lambda$=0.07 and $K$=2.515 were selected. Following recommendations by Wu et al. (2021), σ=0.125 was also used. Figure 5 presents the results, where the consecutive out-of-control events in the original time scale are indicated to facilitate the discussion.

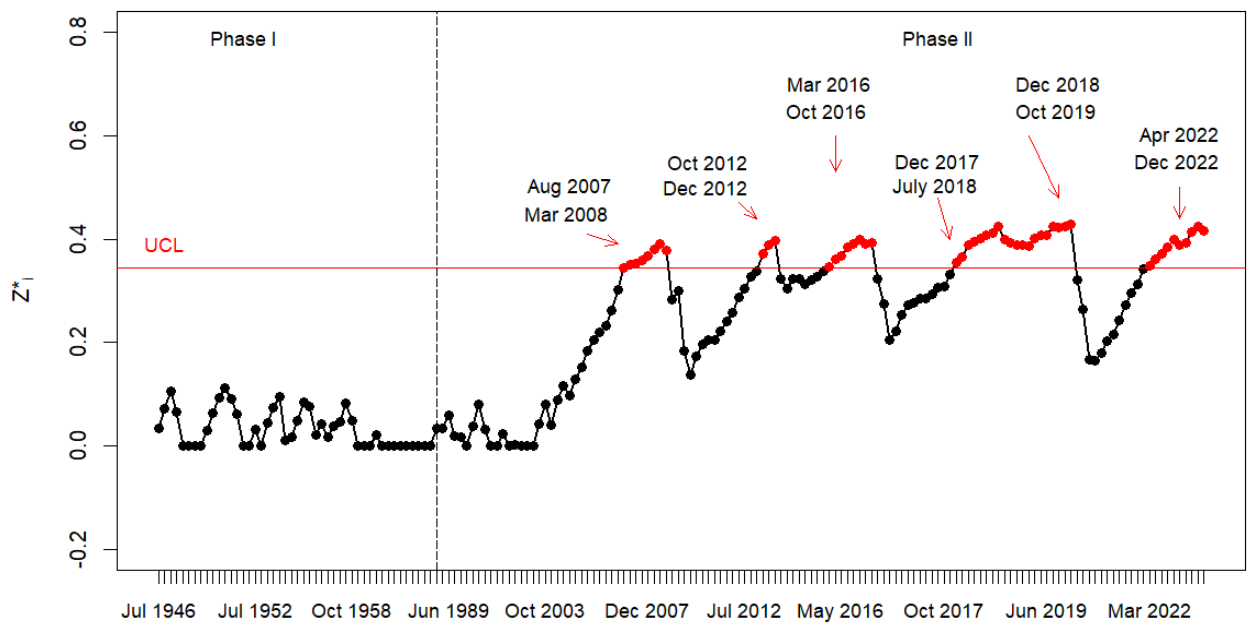

**Figure 5**: results of the EWMA TBEA control chart

## 3.2 The CP control charts

Figure 6 displays results obtained with the Kolmogorov-Smirnov control chart. Please note that adjustments were made to the output of the *cpm* package to better suit our purposes. Figure 6 reports the estimated change points, the statistic $Z_R$ and the averages of $Z_R$ between every pair of change points (dashed line). In this case, four change points were estimated.

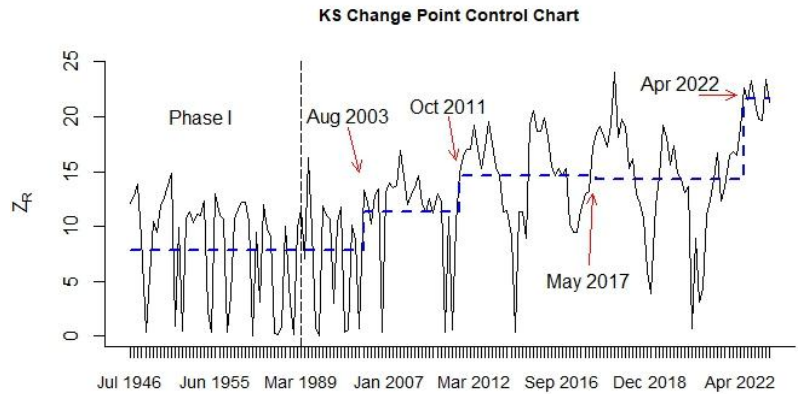

**Figure 6**: KS control chart

Figure 7 illustrates the results obtained with the Mann-Whitney CP control chart, which identified three change points.

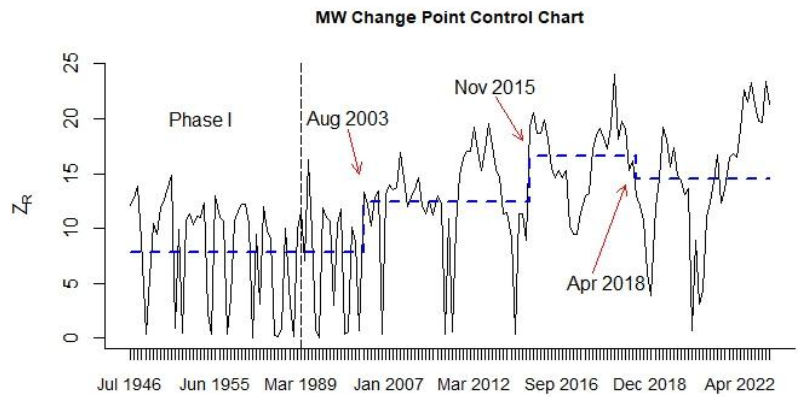

**Figure 7**: Mann-Whitney control chart

## 4 Open Issues and Discussion

Before discussing the results, some important issues need be addressed.

To begin with, no alarms were detected in the Phase 1 dataset. Thus, the estimated medians $\hat{\theta}_{T_0} = 1$, $\hat{\theta}_{X_0} = 1.2263$ and averages $\hat{\mu}_{T_0} = 10.9149$, $\hat{\mu}_{X_0} = 1.2511$ can be considered reliable estimates of the corresponding unknown in-control parameters.

Next, it is important to remember that the EWMA control chart is tailored for monitoring TBEA data, with well-documented statistical properties (Wu et al., 2021). In contrast, the change point control charts are not specifically designed for monitoring time-between-events and amplitude statistics. Therefore, a potential concern is that their false alarm rates may differ from the theoretical values. To assess the performance of the change point control charts, we conducted a simulation study in which, to make the comparisons as consistent as possible, we generate data similar to those observed in the present case.

In Section 3, Phase-1 was performed on 517 SPEI-12 observations, with mean and the standard deviation $\mu_{SPEI-12} = 0.2082$ and $\sigma_{SPEI-12} = 0.9213$, respectively. In these observations occurred 47 events $E$. To reproduce a similar scenario, 500 observations were generated, following a normal distribution with mean $\mu_{SPEI-12}$ and standard deviation $\sigma_{SPEI-12}$. Events ($E$) were identified, and their corresponding magnitudes ($X_i$) and time between two events ($T_i$) were determined. Subsequently, the mean values of $X_i$ and $T_i$ were estimated to configure the KS and MW change point control charts. Once set up the KS and MW control charts, 1000 additional normally distributed observations with mean $\mu_{SPEI-12}$ and standard deviation $\sigma_{SPEI-12}$ were generated. On these observations the events $E$ were identified as the corresponding amplitudes $X_i$ and time between events $T_i$. The values of the statistic $Z_R$ were computed and the possible change points estimated. This procedure was repeated $10^6$ and the false alarm rates were calculated by dividing the total number of estimated change points by the total number of events $E$ identified. Table 2 summarizes the results showing the empirical false alarm rates $\hat{\alpha}_0$.

|  | KS | MW |
|---|---|---|
| $\hat{\alpha}_0$ | 0.00222 | 0.00182 |

**Table 2**. empirical false alarm rates $\hat{\alpha}_0$ obtained from the simulation experiment

The results indicate that the empirical false alarm rates do not exceed the theoretical value and that the KS chart's empirical rate closely matches the theoretical value, warranting greater focus on this control chart in subsequent analyses. The analysis on the performance of the examined control charts is completed in Appendix A, where a further simulation study is carried out to assess their statistical properties in the out-of-control case.

We now turn to discussing the results. In this work, different methodologies were employed. However, the examined control charts consistently suggest worsening drought conditions over time. The TBEA statistics show significant increases beginning in August 2003 (MW and KS control charts, Figures 6 and 7) and August 2007 (EWMA control chart, Figure 5).

Consistency among results becomes evident when the change points identified by the KS control chart are superimposed on the EWMA chart (the continuous vertical lines in Figure 8). The KS control chart frequently estimates change points close to the lower turning points (August 2003, October 2011, July 2015, January 2017 and June 2021 indicated by arrows in Figure 8) of the EWMA statistic.

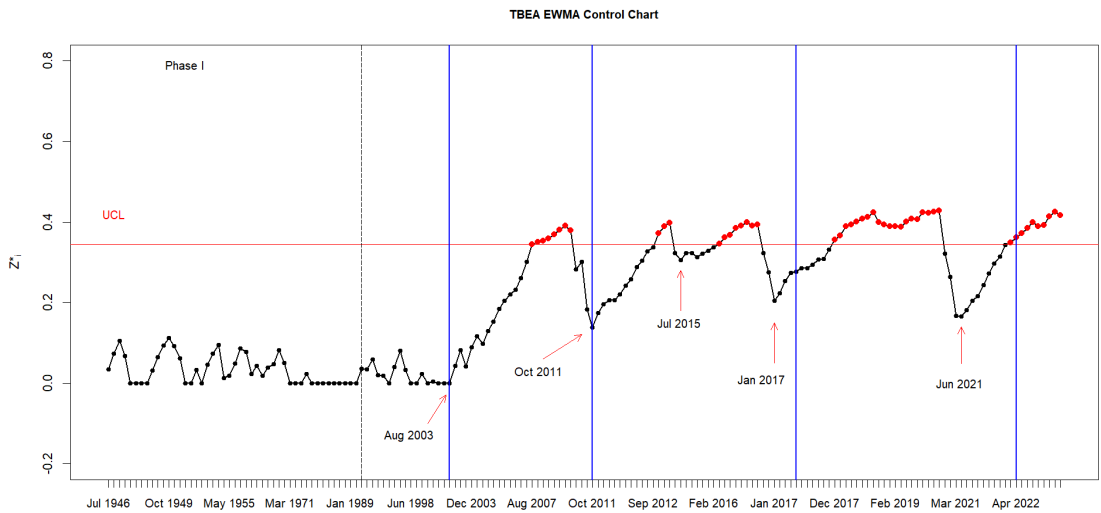

**Figure 8**: EWMA control chart with the change points estimated by the KS control chart (continuous vertical lines) and lower turning points of the EWMA statistic (indicated by arrows)

The amount of agreement in the results can be objectively quantified by performing an Event Coincidence Analysis (ECA; Donges et al., 2016, Siegmund et al., 2017). ECA evaluates empirical co-occurrence frequencies of distinct event types against those expected under the assumption of uncoupled Poisson processes.

In the present case, there are two sequences of events of distinct types:

- events A (the estimated change points) that occur at times $t_i^A$ $(i=1,...,N_A)$;
- events B (the lower turning points of the EWMA statistics) that occur at times $t_j^B$ $(j=1,...,N_B)$.

A coincidence occurs if two events at $t_i^A$, $t_j^B$ with $t_j^B < t_i^A$ are closer in time than a temporal tolerance or coincidence interval $\Delta T$, i.e.

$$t_i^A - t_j^B \leq \Delta T. \tag{15}$$

To quantify the coincidence intensity between Type A and Type B event sequences, we computed the precursor coincidence rate (16):

$$r_p(\Delta T, \tau) = \frac{1}{N_A} \sum_{i=1}^{N_A} \Theta\left[\sum_{j=1}^{N_B} 1_{[0,\Delta T]}\left((t_i^A - \tau) - t_j^B\right)\right] \tag{16}$$

where $\Theta(x) = 0$ for $x \leq 0$ and $\Theta(x) = 1$ otherwise, $1_I(\cdot)$ is the indicator function of the interval $I$ ($1_I(x) = 1$ for $x \in I$ and $1_I(x) = 0$ otherwise) and $\tau \geq 0$ is a time lag parameter to adjust for possible lagged relationships. The precursor coincidence rate, $r_p(\Delta T, \tau)$, measures the fraction of A-type events that are preceded by at least one B-type event within the tolerance interval $\Delta T$.

It is also possible to test the null hypothesis that the observed number of coincidences can be explained by two independent series of randomly Poisson-distributed events. Table 3 summarizes the results for $\Delta T$ ranging from 0 to 5. To consider only the instantaneous coincidences within the coincidence interval $\Delta T$ the results of Table 3 were obtained by setting $\tau = 0$.

| $\Delta T$ | $r_p(\Delta T, \tau)$ | p-value |
|---|---|---|
| 0 | 0.50 | 0.0000000 |
| 1 | 0.50 | 0.0066222 |
| 2 | 0.50 | 0.0245981 |
| 3 | 0.50 | 0.0514191 |
| 4 | 0.75 | 0.0079741 |
| 5 | 0.75 | 0.0145453 |

**Table 3**: ECA for EWMA lower tuning points (events B) and KS estimated change points (events A)

The analysis indicates that even in the most stringent case (i.e. $\Delta T = 0$, meaning coincidence within the same month), the hypothesis that these coincidences occur by chance can be rejected (*p-value*=0.000).

This result confirms a significant degree of consistency between the two methodologies, that can be effectively combined to provide complementary insights. For the EWMA chart, post-signal analyses can be conducted to identify consecutive out-of-control events in the original time scale. For instance, a long period of consecutive out-of-control $Z_i^*$ values occurred from December 2018 to October 2019. For the change point control charts, it is important to recall that we monitor $Z_R = T'/X'$, which simultaneously accounts for both time between drought events and their magnitudes. This means that the estimation of a change point in the sequence of the TBEA statistic correspond to a change in drought characteristics. For example, the KS control chart reveals that increases in the TBEA statistic observed in August 2003 and October 2011 indicate an increase in drought severity. Furthermore, having an estimate of the change point time in a TBEA statistic might allow for a better understanding of the monitored process and the optimization of response strategies.

## 5. The SPEI-24 case

Before moving on to the conclusions, it may be worthwhile to examine the results obtained when the previously introduced methodologies are applied to the SPEI-24 data. The 24-month accumulation period is commonly used to quantify socioeconomic droughts, where prolonged imbalances between water supply and demand result in shortages of essential goods and services. The impact extends beyond the agricultural sector, affecting energy production, sanitation, and industrial processes. As such, monitoring the characteristics of socioeconomic drought events is of considerable importance. The time series plot of the SPEI-24 data for Bologna is displayed in Figure 9. In this case data start from December, 1902 and are available up to December, 2022 (1441 observations).

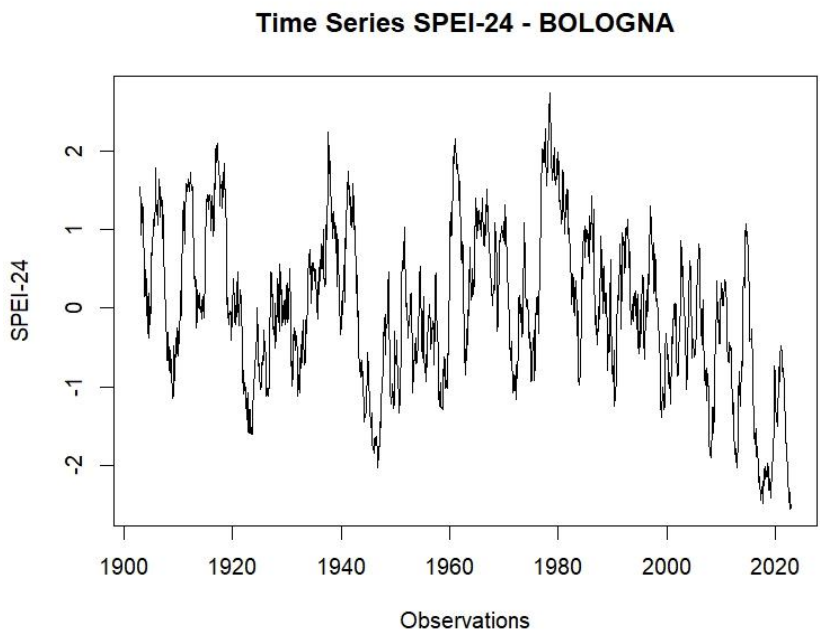

**Figure 9:** Time sequence plot of the available SPEI-24 data for the period December, 1902–December, 2022

In total there are 219 Events $E$ (values of SPEI-24≤-1) and Figures 10 and 11 illustrate the absolute values SPEI-24≤-1 ($X_i$) and the time ($T_i$) between the drought events $E$, respectively.

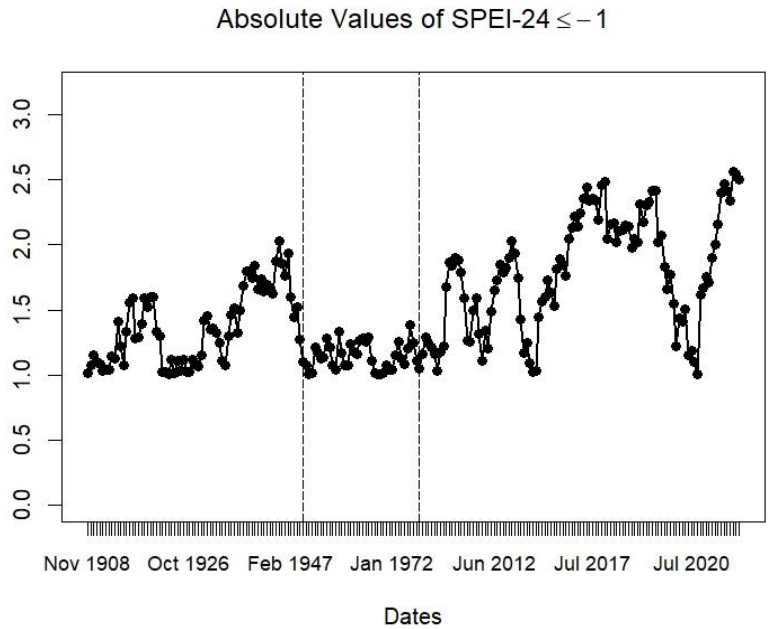

**Figure 10**: Amplitudes $X_i$ of the drought events $E$ (data for Phase-1 are indicated between dashed vertical lines).

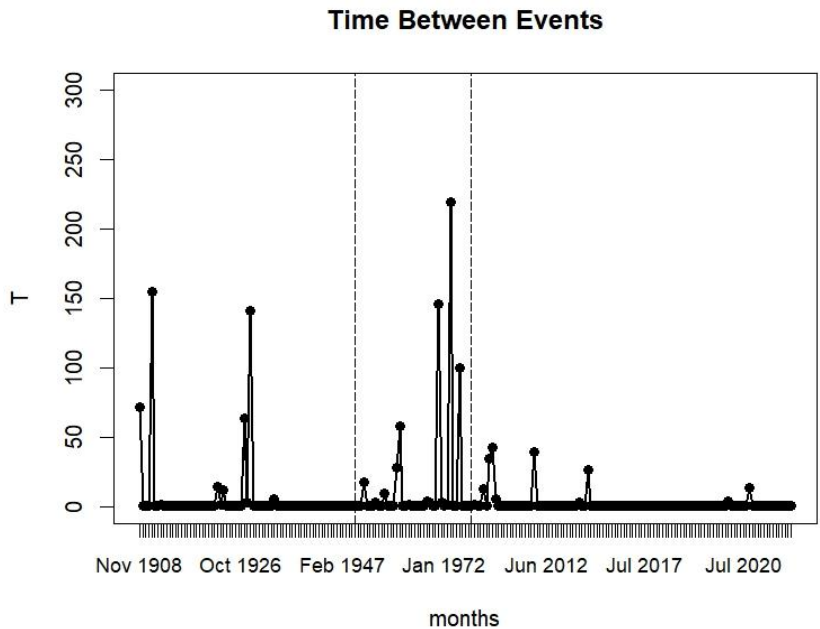

**Figure 11** Time ($T_i$) in months between drought events $E$ (data for Phase-1 are indicated between dashed vertical lines) for the SPEI-24 case.

For performing the Phase-1 we selected the time period from June 1947 to April 1999, comprising 623 SPEI-24 observations, with mean and the standard deviation $\mu_{SPEI-24} = 0.2522$ and $\sigma_{SPEI-24} = 0.8604$, respectively. In this period, marked by dashed vertical lines in Figures 10 and 11, occurred 40 events $E$ (values of *SPEI*-24≤-1), which were used to estimate in-control median (for the EWMA chart) and mean (for the CP charts) values: $\hat{\theta}_{T_0} = 1$, $\hat{\theta}_{X_0} = 1.1285$, $\hat{\mu}_{T_0} = 15.575$ and $\hat{\mu}_{X_0} = 1.1428$, respectively. Also in this case, to ensure comparability all control charts were configured with the same in-control theoretical statistical properties (ARL$_0$=370 or with a theoretical false alarms rate equal to $\alpha_0 = 0.0027$).

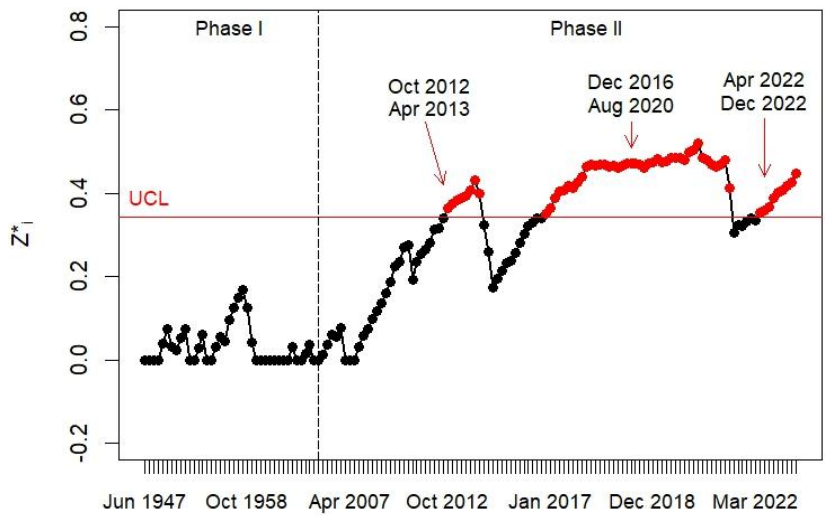

**Figure 12**: results for the EWMA TBEA control chart (SPEI-24 case)

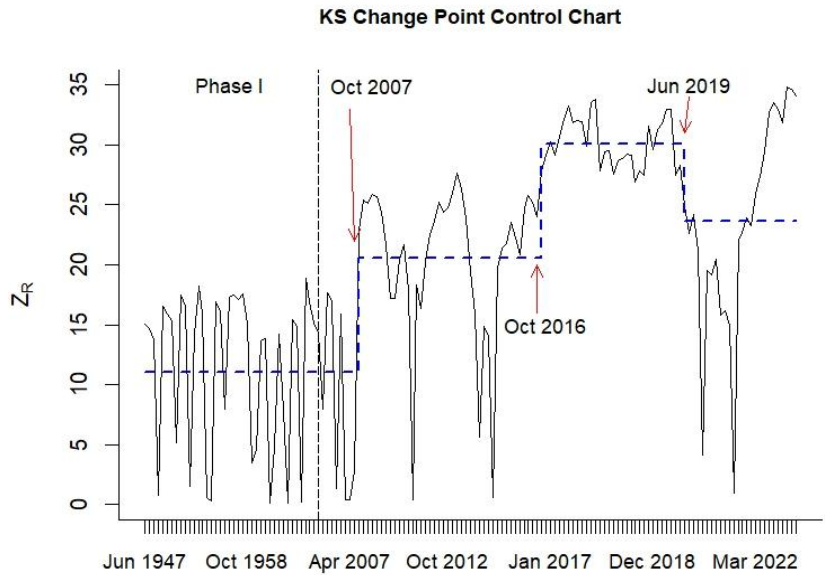

**Figure 13**: results for the KS control chart (SPEI-24 case)

In this case, according to the SPEI-24 data, drought conditions begin to deteriorate starting in October 2007 (Figure 13) and the EWMA control chart identifies a first drought period from October 2012 to April 2013 (Figure 12).

The consistency between the results obtained from the two control charts is illustrated in Figure 14, where the averages -appropriately rescaled- of the $Z_R$ statistics between each pair of change points are superimposed on the EWMA statistic.

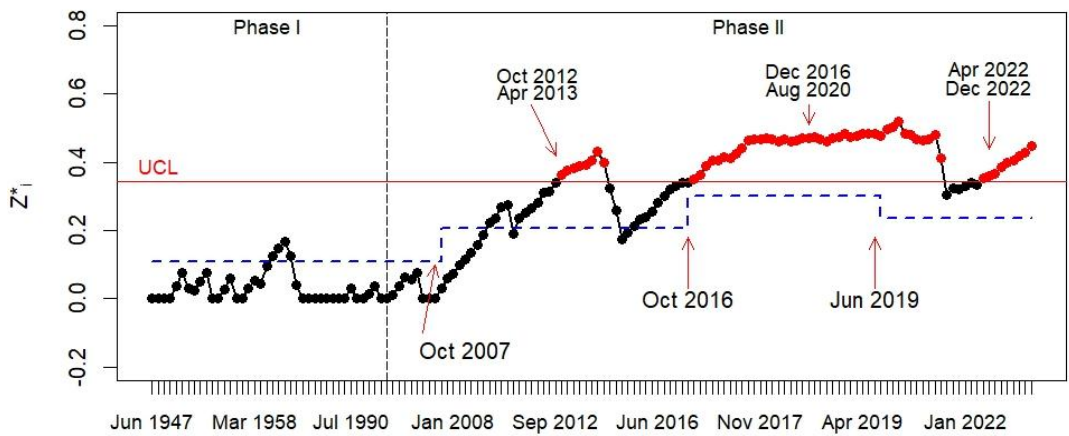

**Figure 14**: EWMA control chart (SPEI-24 case) with the averages (dashed line and appropriately rescaled) of $Z_R$ statistic between every pair of change points.

In this Figure, it can be observed that the EWMA statistic exhibits an upward trend starting in October 2007, which is correctly identified by the KS control chart. Further analysis of Figure 14 reveals a second, significantly longer dry period, extending from December 2016 to August 2020, detected by the EWMA control chart. Notably, the TBEA trend begins earlier, while the KS chart identifies the corresponding change point (October 2016) with a delay relative to the EWMA statistic's lower turning point. Nevertheless, the KS change point control chart successfully captures the intensification of drought episodes, in agreement with the findings from the EWMA chart. The most recent change detected by the KS chart, in June 2019, suggests a slight improvement in drought conditions - a decreasing trend also mirrored in the EWMA statistics.

## 6. Concluding Remarks

In the context of climate change, the frequency, intensity, and duration of droughts are expected to rise, making it crucial to monitor these characteristics effectively.

In this article, we studied the contribution of distribution-free TBEA control chart to detect changes in the characteristics of drought events. A distribution-free upper sided EWMA-TBEA control chart was considered, along with the proposal to use non-parametric change-point control charts.

As the change-point control charts were not specifically designed for TBEA monitoring, we examined their statistical properties through a simulation study. Furthermore, we assessed the coherence of the obtained results.

The methodologies examined and proposed in this article provided consistent and useful results that can be used in a complementary way:

• With the EWMA control chart, it is possible to perform a post-signal analysis to highlight out-of-control drought episodes that are consecutive with respect to the original time scale;

• With the change point control charts, it is possible to identify significant changes in drought characteristics.

To summarize, TBEA control charts may provide a valuable contribution to institutions responsible for developing strategies to mitigate and manage the adverse effects of drought.

In any future work it would be desirable to extend this analysis several directions. This study exclusively considered data from one geographical location. Therefore, it would be useful to extend this work to data from other geographical areas. Second, it would be interesting to consider change points control charts based on different statistics such as Lepage and Cramer-von-Mises control charts (Ross 2015).

# Appendix A

To study the statistical properties of the control charts in the out-of- control scenario, for each run, of the $k=10^6$ replications, we generate $n=1500$ observations: the first $m$ observations are generated from a Gaussian distribution with mean $\mu_{SPEI-12}$ and standard deviation $\sigma_{SPEI-12}$, while the observations from $m+1$ to $n$ are generated are generated from a Gaussian distribution with mean $\mu_1 = \mu_{SPEI-12} + \delta \times \sigma_{SPEI-12}$ and standard deviation $\sigma_{SPEI-12}$.

In the first in-control $m$ observations, events ($E$) were identified, and their corresponding magnitudes ($X_i$) and time between two events ($T_i$) were determined. Subsequently, the mean and median values of $X_i$ and $T_i$ were estimated to configure the EWMA, KS and MW control charts. Once set up the control charts, on the 1000 out-of-control observations $E$ were identified as the corresponding amplitudes $X_i$ and time between events $T_i$. The values of the statistics $Z^*$ and $Z_R$ were computed and using the EWMA, KS and MW control charts the possible alarms were detected and change points estimated. The number of observations taken from the point of change $m$ till the alarm instant is the out-of-control run length (RL). The average of the $k$ run lengths is our estimate of the out-of-control average run length (ARL$_1$). In each run, if at the end of the $n$ observations the control chart did not detect any alarm, we defined this event as a missed alarm. Therefore, we compute for each control chart the percentage of missed alarms obtained in the $k$ replications. More specifically, as out-of-control cases, we considered the $\delta = (-0.5, -1, -1.5)$.

The simulation results are summarized in Tables A1–A3 where for each control chart and each simulated scenario the following quantities are reported: %missAL, the percentage of missed alarms; ARL1, the estimated average length; SDRL, the standard deviation of the RL.

|  | EWMA | KS | MW |
|---|---|---|---|
| **%missAL** | 0.3107 | 1.6447 | 1.4524 |
| **ARL$_1$** | 23.3798 | 24.5281 | 26.1915 |
| **SD** | 21.9405 | 26.1601 | 25.2083 |

**Table A1**. Simulation results for $\delta$=-0.5

|  | EWMA | KS | MW |
|---|---|---|---|
| **%missAL** | 0 | 0.2692 | 0.1696 |
| **ARL$_1$** | 9.8492 | 9.1902 | 9.0464 |
| **SD** | 4.9645 | 18.1281 | 11.5032 |

**Table A2**. Simulation results for $\delta$=-1

|  | EWMA | KS | MW |
|---|---|---|---|
| **%missAL** | 0.0000 | 0.0588 | 0.0404 |
| **ARL$_1$** | 7.5120 | 6.6385 | 6.4334 |
| **SD** | 2.9521 | 18.2150 | 10.0409 |

**Table A3**. Simulation results for δ=-1.5

The results reported in Table A1-A3 show that in terms of ARL$_1$ the control charts are comparable: the EWMA chart has a slightly lower percentage of undetected alarms than the others, but all things considered the control charts have similar performance.